# Multi-Task Mixture Density Graph Neural Networks for Predicting Cu-based Single-Atom Alloy Catalysts for CO$_2$ Reduction Reaction

Chen Liang [a, 1], Bowen Wang [b, 1], Shaogang Hao [c], Guangyong Chen [d]*, Pheng-Ann Heng [b], Xiaolong Zou [a]*

a. Shenzhen Geim Graphene Center, Tsinghua-Berkeley Shenzhen Institute & Tsinghua Shenzhen International Graduate School, Tsinghua University, Shenzhen 518055, China

b. Department of Computer Science and Engineering, The Chinese University of Hong Kong, Hong Kong 999077, China

c. Tencent, Shenzhen 518054, China

d. Zhejiang Lab, Zhejiang University, Hangzhou 311121, China

1 These authors contributed equally to this work.

* Corresponding authors: gychen@zhejianglab.com; xlzou@sz.tsinghua.edu.cn

# Abstract

Graph neural networks (GNNs) have drawn more and more attention from material scientists and demonstrated a high capacity to establish connections between the structure and properties. However, with only unrelaxed structures provided as input, few GNN models can predict the thermodynamic properties of relaxed configurations with an acceptable level of error. In this work, we develop a multi-task (MT) architecture based on DimeNet++ and mixture density networks to improve the performance of such task. Taking CO adsorption on Cu-based single-atom alloy catalysts as an illustration, we show that our method can reliably estimate CO adsorption energy with a mean absolute error of 0.087 eV from the initial CO adsorption structures without costly first-principles calculations. Further, compared to other state-of-the-art GNN methods, our model exhibits improved generalization ability when predicting catalytic performance of out-of-domain configurations, built with either unseen substrate surfaces or doping species. We show that the proposed MT GNN strategy can facilitate catalyst discovery.

**Keywords:** Machine learning; $CO_2$ reduction reaction; Graph neural network; Multi-task learning; Mixture density network

# Introduction

CO2 reduction reaction (CO2RR), which plays an important role in alleviating environmental problems caused by the greenhouse effect [1-3], has aroused a widespread research interest. Numerous attempts have been made to comprehend the underlying mechanism of CO2RR [4-8], and it has been found that the adsorption energy of CO ($E_{CO}$) acts as a good descriptor for the CO2RR performance of catalysts [9-11]. The vast choice of different catalysts renders the exploration of full configuration space of CO adsorption impractical. Fortunately, machine learning (ML) [12] has demonstrated a promising potential for accelerating the process of the design and optimization of catalysts, based on experimental and computational datasets [13-16]. Until now, many ML models have been proposed to predict $E_{CO}$ in order to guide the design of high-performance CO2RR catalysts, including ensemble learning algorithms based on decision trees [17-21] and artificial neural networks [22, 23].

In contrast to traditional feature-based algorithms, graph neural networks (GNNs) [24-30] have shown better performance and good interpretability [31-33], due to the analogy between the constructed data structure and real geometric configurations [34] and the introduction of message-passing processes [35]. The early application of GNNs to the prediction of chemical properties [36] demonstrates superior performance in cheminformatics compared to conventional methods [37]. Following this, various GNN models, such as DTNN [38], SchNet [39], HIP-NN [40], CGCNN [41], MEGNet [42], PhysNet [43], AttentiveFP [44], DimeNet [45, 46], PAINN [47], GemNet [48], HermNet [49], etc., are proposed with the prediction accuracy continuously improving. However, few models can predict the target adsorption energies for relaxed structures directly from initial unrelaxed structures with an acceptable accuracy [50, 51]. Besides, these models show poor performance when extrapolating to untrained configurations, severely limiting their applicability.

In this work, we develop a new GNN model based on DimeNet++ [46] to discover Cu-based single-atom alloy (SAA) catalysts, which are among the most promising catalysts for CO2RR [52-57], significantly accelerating the catalyst optimization procedure by circumventing first-principles calculations. We aim to use initial unrelaxed structures as input to accurately estimate $E_{CO}$ on the corresponding relaxed surfaces by a multi-task (MT) architecture [58, 59] with two branches for two distinct but interconnected tasks. Mixture density networks (MDNs) [60-62] are further integrated into one of the branches to incorporate relaxed structure information in the training process, while the information is no longer needed in the inference process. The proposed model outperforms representative benchmark GNNs and conventional algorithms on a dataset containing 3075 different Cu-based SAAs adsorbed by CO molecules with the adsorption energies calculated based on density functional theory (DFT), and a good extrapolation performance is also shown on catalyst configurations built with either unseen substrate surfaces or doping species. In addition, the significance of MDNs is examined in our experiments.

# Results

### Multi-task mixture density DimeNet++

Directional Message Passing Neural Network [45], or DimeNet for short, is one of the state-of-the-art GNN methods for predicting the properties of molecules and materials. By default, DimeNet is comprised of one Embedding Block (EB), constructing initial directional edge features (or embeddings) based on the node features (or embeddings), and six Interaction Blocks (IBs), performing updating and aggregation of edge embeddings. DimeNet extends the concept of message-passing from node embeddings to directional edge embeddings, therefore enhancing the ability to differentiate among various molecules. As an improved version of DimeNet, DimeNet++ [46] drastically decreases the training time needed while maintaining high accuracy by replacing the bilinear layer with element-wise multiplication and adding more multilayer perceptrons (MLPs). Additionally, the number of IBs used in the algorithm has also been reduced from six to four.

Our proposed model's architecture, named multi-task mixture density (MT-MD) DimeNet++, is illustrated on the right panel of **Fig. 1a**, with DimeNet++ shown on the left for comparison. The model is designed under the MT learning framework, comprised of two branches for two distinct but related objectives. The main branch (MB) is used to predict adsorption energy, while the structure relaxation branch (SRB) is used to predict the change of atomic positions after relaxation in the form of bond length change. It is reasonable to assume that the change of pair-wise distance among atoms follows a distribution rather than a set of fixed values, and MDNs [60], a type of specially designed artificial neural networks for approximating conditional distribution of labels, are applied in SRB. Additionally, MB and SRB are separated into two branches at the beginning of the architecture, and cross-stitch units

(CSUs) [63], a state-of-the-art MT learning technique, are applied to build a connection between the two branches in the network, in order to determine the ideal network architecture for each task to prevent the possible negative transfer.

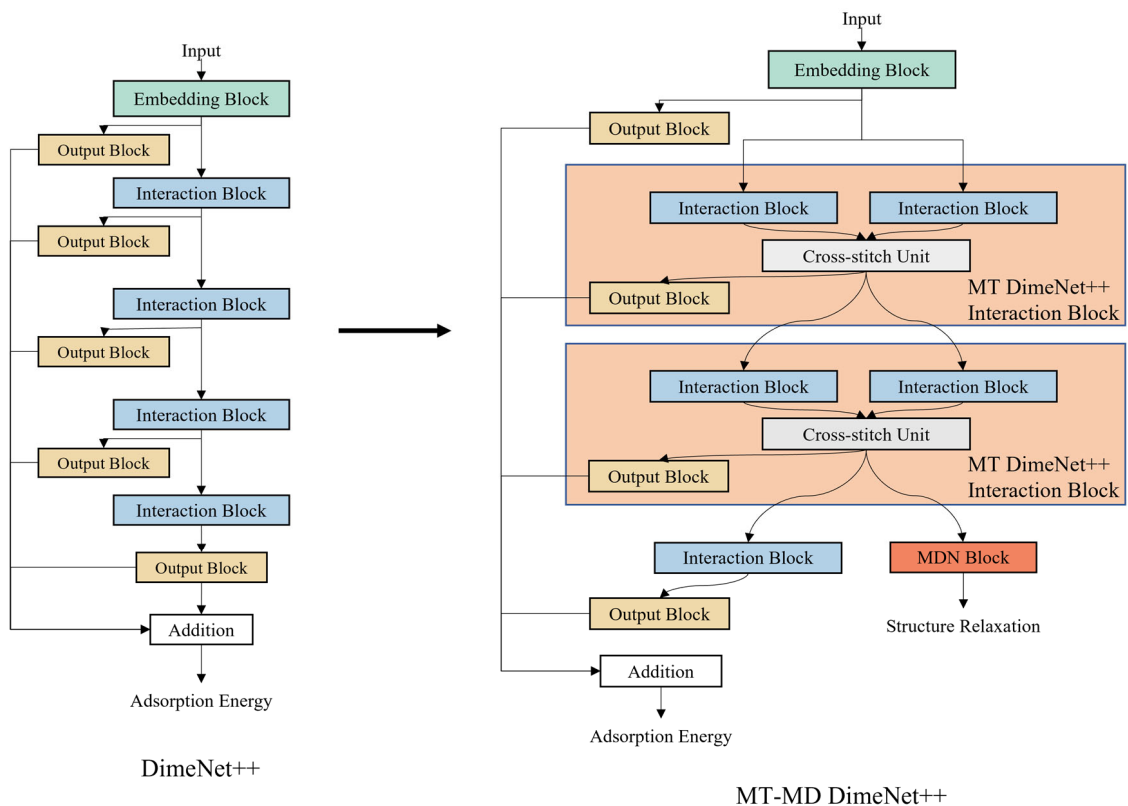

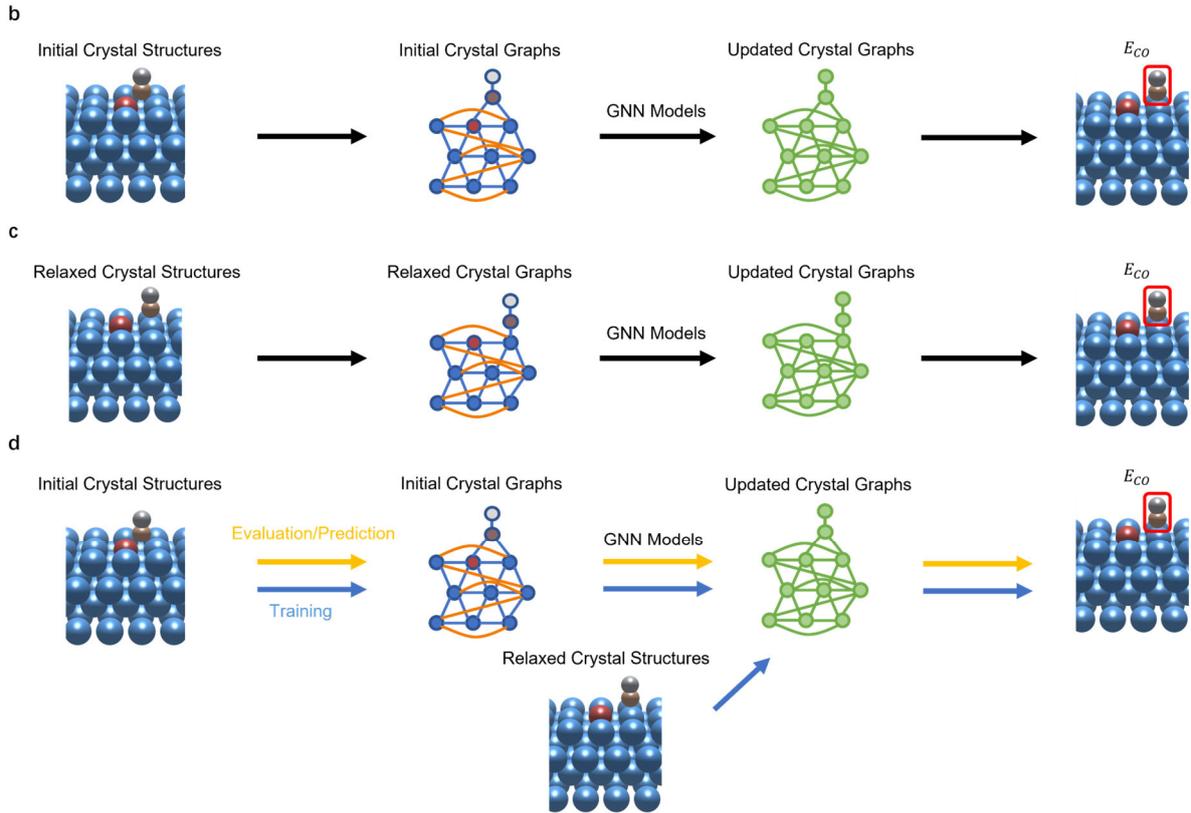

**Figure 1 | The architecture of multi-task mixture density DimeNet++ and the workflows of the prediction tasks. (a)** The structures of DimeNet++ (left) and its multi-task variant (right). Note that we re-organize the plot of operations in IB from their original form [46] in the way that the Output Block inside IB has been taken out, in order to make the illustration clearer. **(b-c)** The workflow of predicting $E_{CO}$ with initial or relaxed crystal structures. Based on the cut-off radius and maximum number of neighbors, the input crystal structures are encoded into crystal graphs, in which the silver, brown, red, and blue nodes represent the O, C, dopant, and substrate atoms, respectively. The additional orange edges are built to satisfy the periodic boundary condition. **(d)** The MT workflow proposed in this work. The blue arrows denote the training process, and the yellow arrows denote the inference process.

**Benchmarks on basic adsorption energy prediction tasks**

A dataset containing 3075 Cu-based SAAs and corresponding $E_{CO}$ [20] is applied in this work, in which 41 different element species are selected as dopants. By varying the surface index, the doping position, and the adsorption site of the CO molecule, 75 distinct structures are constructed for each doping species, composed of 6, 8, 8, 27, and 26 structures on the Cu(100), Cu(110), Cu(111), Cu(210), and Cu(411) surfaces, respectively. The whole dataset is split 10 times randomly with 60% being the training set, 20% being the validation set, and the remaining 20% being the test set, respectively. Every model in the tests in this section is trained and evaluated on the same 10 splits of the dataset to make a fair comparison.

Two fundamental tasks are depicted in **Figs. 1b** and **1c**, employing initial and relaxed structures as inputs to GNNs to predict $E_{CO}$ obtained after optimization, respectively, in which the initial structures are defined as the unrelaxed ones. The two tasks are comparable to the IS2RE task and S2EF task defined in Open Catalyst Project (OCP) [64], excluding force. We refer the former as the "I2I" task because both the training and test sets contain initial structures, whereas the latter is referred to as the "R2R" task because both the training and test sets contain relaxed structures. In addition, we establish the "R2I" task, where models are trained to predict adsorption energies using relaxed structures as inputs, but use initial structures to predict corresponding energies during inference. The labels used in all three tasks are energies derived after DFT structural relaxation.

The dataset is benchmarked by performing the three tasks with several representative baseline models. **Table 1** presents a comprehensive comparison of the 8 models on I2I and R2R tasks, in which AEVR. and STD. denote the average and standard deviation of evaluation metrics on 10 test runs, respectively. The STDAE means the standard deviation of absolute errors of test results, and a high STDAE implies that some energies estimated by the model depart significantly from their actual values calculated by DFT, indicating a weak generalization ability. The metrics "within 0.02" and "within 0.1" represent the ratio of samples in test sets predicted with an absolute error smaller than 0.02 and 0.1 eV, respectively. Three symmetry-function (SF) [65, 66] based Gradient Boosting Regression (GBR) [67] models are listed as representatives of traditional ML algorithms, with doping and carbon atoms chosen as centers to create the SFs. Details of constructing SFs and the hyper-parameters of GNNs are available in **Supplementary Material**. The results show that all models have a higher accuracy when predicting the adsorption energy with relaxed structures instead of their initial counterparts, which is consistent with earlier works [50, 64]. GBR models have a substantially lower average MAE than SchNet and CGCNN on both I2I and R2R tasks, but their STDAE values are larger than those of GNN models, indicating that GBR models can only perform well on some of the data points. In comparison, DimeNet-based models perform the best on all metrics, though the run time and the number of trainable parameters exceed those of SchNet and CGCNN. In R2R task, DimeNet is superior to DimeNet++ with the original hyper-parameters (embedding size = 128, output embedding size = 256), but the performance of DimeNet++ with the embedding size = 256 and output embedding size = 192 (denoted as

DimeNet++ (emb256)) is identical with or slightly better than that of DimeNet. Hereafter, the DimeNet++ (emb256) is applied for further experiments.

**Table 1. Comparison of models on I2I, MT I2I and R2R tasks.**

| Task | Model | Test MAE AEVR. (eV) | Test MAE STD. (eV) | Test STDAE AEVR. (eV) | Test within 0.02 AVER. | Test within 0.1 AVER. | Run Time AVER. (s) | Number of Parameters |
|---|---|---|---|---|---|---|---|---|
| I2I | Localized cos [66] + GBR | 0.120 | 0.0178 | 0.457 | 0.299 | 0.758 | 6.83 | \ |
| | Gaussian-cos [65] + GBR | 0.130 | 0.0192 | 0.482 | 0.271 | 0.738 | 6.82 | \ |
| | Gaussian-tanh [65] + GBR | 0.132 | 0.0201 | 0.485 | 0.268 | 0.735 | 6.71 | \ |
| | SchNet [39] | 0.257 | 0.113 | 0.421 | 0.103 | 0.409 | 1845.45 | 541697 |
| | CGCNN [41] | 0.182 | 0.0887 | 0.401 | 0.147 | 0.557 | 1712.22 | 703361 |
| | DimeNet [45] | 0.099 | 0.0157 | 0.358 | 0.329 | 0.792 | 7741.72 | 2100070 |
| | Dimenet++ [46] | 0.0966 | 0.0146 | 0.350 | 0.312 | 0.802 | 5335.63 | 1887110 |
| | DimeNet++ (emb256) | **0.0938** | 0.0148 | 0.356 | 0.337 | 0.821 | 5906.27 | 3545542 |
| MT I2I | MT-MD DimeNet++ (MB3-SRB2) | 0.0875 | 0.0135 | 0.310 | 0.384 | 0.823 | 7258.57 | 4569855 |
| | MT-MD DimeNet++ (MB4-SRB3) | **0.0870** | 0.0114 | 0.303 | 0.398 | 0.829 | 9231.51 | 6250431 |
| R2R | Localized cos [66] + GBR | 0.116 | 0.0237 | 0.585 | 0.311 | 0.792 | 26.25 | \ |
| | Gaussian-cos [65] + GBR | 0.108 | 0.0138 | 0.594 | 0.361 | 0.800 | 25.42 | \ |
| | Gaussian-tanh [65] + GBR | 0.117 | 0.0122 | 0.643 | 0.358 | 0.801 | 26.03 | \ |
| | SchNet [39] | 0.183 | 0.150 | 0.386 | 0.201 | 0.628 | 1904.08 | 541697 |
| | CGCNN [41] | 0.139 | 0.102 | 0.355 | 0.198 | 0.683 | 1780.57 | 703361 |
| | DimeNet [45] | 0.0556 | 0.0138 | 0.312 | 0.569 | 0.908 | 7678.62 | 2100070 |
| | Dimenet++ [46] | 0.0603 | 0.0116 | 0.315 | 0.466 | 0.898 | 5171.22 | 1887110 |
| | DimeNet++ (emb256) | **0.0551** | 0.0137 | 0.312 | 0.544 | 0.911 | 5893.75 | 3545542 |

**Tables S1** and **S2** show the test results achieved by models trained using normalized training labels for the I2I and R2R tasks, respectively. Because the normalization approach could not guarantee a consistent performance increase for all algorithms, the labels of training sets in the following experiments are not normalized during the training phase. Results of R2I task, shown in **Table S3**, exhibit a large MAE for all models. This proves that training models to distinguish new catalysts by their initial structures can be an ineffective way.

**Adsorption energy prediction with MT GNN models**

Most ML models, including GNNs, cannot predict $E_{CO}$ with a high accuracy for I2I task. To solve such a challenge, an MT learning framework is proposed, whose workflow is illustrated in **Fig. 1d**. In these tests, initial structures are applied as the inputs to predict $E_{CO}$, and the information of relaxed structures is included during training by adding a new branch to predict the change of bond length. The trained models are then applied to predict directly $E_{CO}$ using initial structures without relaxation. **Table S4** compares the performance of DimeNet++ and MT-MD DimeNet++ with multiple architectures on the 10 test sets of I2I task, and results of two MT-MD DimeNet++ models are also listed in the "MT I2I" category in **Table 1**. In these tables, the numbers behind 'MB' and 'SRB' indicate the numbers of IBs used in MB and SRB, respectively.

As is shown in **Table S4**, all MT-MD DimeNet++ models outperform original DimeNet++. Notably, even the model utilizing only one IB in MB and SRB (MB1-SRB1), where both the number of trainable parameters and run time are less than those of DimeNet++, has an improved prediction accuracy with an average MAE of 0.0926 eV. These results clearly show the superiority of our method. The best model on the MT I2I task is MB4-SRB3 with an MAE of 0.0870 eV and 9231.51s needed to train for once. To balance the performance and computational costs, MB3-SRB2 with an MAE 0.0875 eV is chosen for further discussion.

**Extrapolation on different surfaces**

It is now well-known that different Cu facets lead to different products for $CO_2$RR [68-72]. More and more experiments are designed to realize a high product selectivity by synthesizing Cu nanostructures with specific facets. Accordingly, a good extrapolation capability on different surfaces is the key to applying ML models to facilitate the optimization of $CO_2$RR catalysts. The demonstration of the generalization capacity of our model is proceeded by training the model to predict $E_{CO}$ of SAAs on untrained surfaces.

In our dataset, there are 246, 326, 328, 1107, and 1066 samples based on Cu(100), Cu(110), Cu(111), Cu(210), and Cu(411) surfaces, respectively. By randomly selecting 246 samples built upon each surface, five subsets are constructed. One of the five subsets is chosen as the target, which is further split half-and-half as the validation and test sets, while the remaining 4 subsets are combined as the training set, leading to a dataset with an 8: 1: 1 split ratio, so that the number of training samples is the same for different surface extrapolation tasks. Based on these datasets, MT-MD DimeNet++ is trained to predict $E_{CO}$ of SAAs which are built on surfaces that do not appear in the training sets. The process is conducted 10 times with the same 10 random seeds for each of the five surfaces, and the average MAEs are depicted in **Fig. 2a**, with detailed results reported in **Table S5**. Although trained on a relatively small dataset (984 samples in total), our model still achieves average MAEs below 0.2 eV on 4 out of 5 test sets. For the Cu(110) surface, we observe that the high MAE is caused by 4 outliers from the Sb-doped case, where the DFT calculated $E_{CO}$ is larger than 10 eV. These apparently unphysical values are attributed to a great reconstruction of the subsurface structure after optimization, shown in **Fig. S1**, which is not the focus of our work. The data points are not removed in previous tests to avoid data leakage.

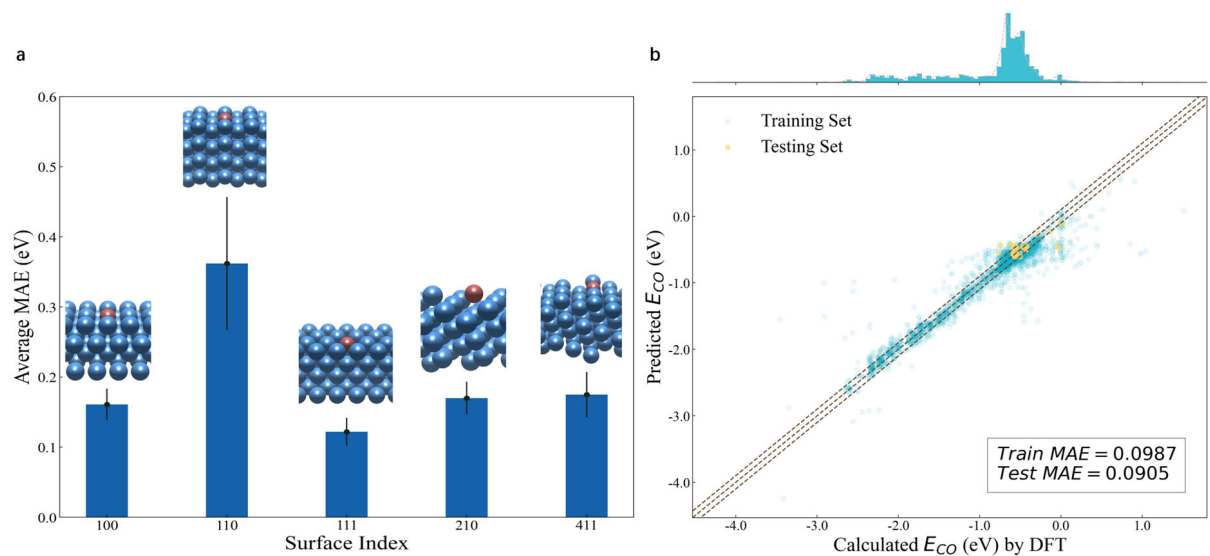

**Figure 2 | Extrapolation tests on different surfaces. (a)** Average test MAEs of extrapolation on 5 different surfaces. Labels on the x-axis denote the indices of the target surfaces, and the configurations of the target surfaces are illustrated above. The black lines denote error bars, reflecting the standard deviation of ten tests. **(b)** The predicted $E_{CO}$ versus their true values calculated by DFT. The cyan points denote the prediction results on the training set and yellow points stand for test results. The brown line in the center is the diagonal, and the lines on both sides are derived by adding and subtracting 0.1 eV, respectively. The distribution of calculated $E_{CO}$ of training samples is exhibited above.

To further demonstrate the extrapolation ability of MT-MD DimeNet++, we evaluate our model by predicting $E_{CO}$ of SAAs based on Cu(751), a surface that has been proved to be more selective for C-C coupling than Cu(100) [73]. We choose 4 doping species (Ag, Zn, Ga and Ge), 2 from d-block and 2 from p-block, which are the most promising doping elements to build Cu-based SAA catalysts for $CO_2RR$ due to their high activity and selectivity, high stability, non-toxicity and low cost [20, 74, 75]. Together with 9 initial structures for each doping case, 36 SAA configurations on Cu(751) are calculated in total to build a new test set. The 4 outliers of Sb-doped Cu(110) mentioned above are removed from the original SAA dataset, and the remaining 3071 samples are applied as the training set. The results in **Fig. 2b** show that both the training and test MAEs are below 0.1 eV, and most predicted values are within the acceptable 0.1 eV error threshold (71.8% for the training set and 69.4% for the test set), reflecting an extraordinary extrapolation capacity of our model.

Additionally, to compare MT-MD DimeNet++ with baseline algorithms, all models are then trained on the dataset, which has been shuffled with 5 different random seeds, to perform the Cu(751) extrapolation test for 5 times, with results shown in **Table S6**. Three DimeNet-based models reach the lowest MAEs compared with other models, and MT-MD DimeNet++ is the only one with an MAE below 0.09 eV (0.0893 eV).

**Extrapolation on different elements**

The doping species have a great impact on the chemical properties of SAAs, leading to different substrate-CO interaction behaviors. The ability to predict $E_{CO}$ of SAAs with unseen doping species is another important aspect to accelerate the development of new SAA catalysts. In this section, each of the 41 doping elements is applied as the target element, respectively, to construct 41 datasets. In each dataset, 75 SAAs doped by the target atom form the test set, while the remaining samples, in which the target element does not appear, form the training set. When Sb is not the target element, the training set contains 2996 data points, excluding 4 Sb-containing samples with $E_{CO}$ larger than 10 eV, whereas the number of training data is 3000 for the case with Sb as the target element. MT-MD DimeNet++ performs the prediction task for 5 times on each target element with 5 different random seeds, and the results are shown in **Fig. 3a** in a periodic table with more details provided in **Table S7**. In total, the average test MAEs of 11 out of 41 elements are within the range of acceptable error (0.10-0.15 eV). The model can predict most untrained non-metal elements with a relatively high accuracy. Although the model performs the best for Cd and Pd (with an MAE of 0.100 and 0.101 eV, respectively), it does not show a good performance when extrapolating to some transition metal elements, especially those in VIB, VIIB groups and the first two columns of VIIIB group. Our preliminary interpretation for these phenomena is that the relatively higher complexity of the interaction pattern between the transition metal elements with Cu substrates benefits the extrapolation on the simpler non-metal doping elements. We leave the study of the underlying mechanism in our future work.

In order to further assess the generalization ability, a series of transfer learning tasks are designed for the three target elements with the worst performance (except Sb), i.e., N, Re and Ru. In this experiment, 2, 5, 10, 20, and 30 data points from the initial test sets are randomly selected and added to the training set to build transfer learning tasks of decreasing levels of difficulty. For each task, the sampling is repeated 5 times based on 5 different random seeds, resulting in 25 datasets in total for each element. All of the baseline models are applied to make a comparison with MT-MD DimeNet++, and the average MAEs for all tasks are plotted in **Figs. 3b-d**, details of which are available in **Tables S8-S10**. Experiments in which no extra data points are supplemented are also conducted for baseline models, and the results are also derived as an average of 5 tests for each model. None of the models can predict $E_{CO}$ of N-doped SAAs accurately, and the three GBR-based models even perform worse when two N samples are included. It is noted many $E_{CO}$ on N-doped samples (23 out of 75) are smaller than -1.0 eV, largely deviated from the mean of the $E_{CO}$ distribution for the whole dataset. This indicates that there is an unusual pattern of interaction with CO for these N-containing catalysts suggested by the reconstructed structures after optimization, leading to the failure of extrapolation. Compared with N-doped cases, test MAEs on Re- and Ru-doped samples show a gradual decrease as the number of Re/Ru samples added increases. For Re-doped SAAs, MT-MD DimeNet++ is the only model achieving an average MAE lower than 0.1 eV (0.0881 eV) when only 10 Re samples are trained, and it continues to show the best performance till 30 extra samples are included (0.0739 eV), notably better than DimeNet++ (0.0895 eV). For Ru doped SAAs, the transfer is relatively easy compared with the other two cases. DimeNet++ (0.0533 eV) outperforms MT-MD DimeNet++ (0.0657 eV) for datasets supplemented with 10 Ru samples, while in the task with 30 Ru samples, the MT-MD one is still better than DimeNet++ with an MAE of 0.0310 eV versus 0.0356 eV. Therefore, although MT-MD DimeNet++ cannot directly predict $E_{CO}$ of SAAs doped with untrained species, the model reaches a high accuracy as long as around 5 to 10 extra samples are included as part of the training samples, largely reducing the computational costs.

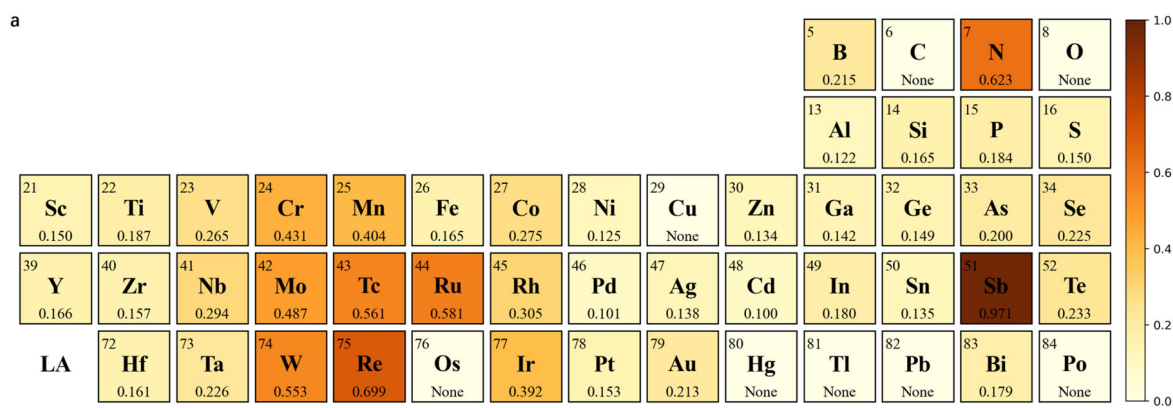

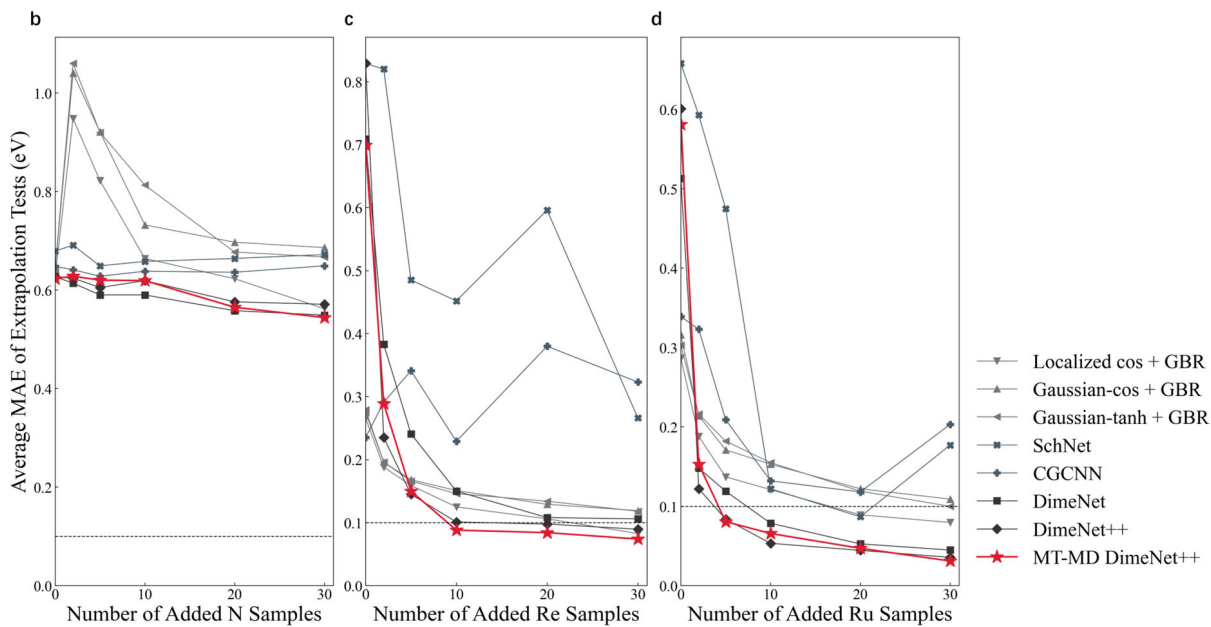

**Figure 3 | Extrapolation tests on different elements. (a)** Average MAEs map (in eV) for MT-MD DimeNet++ on element extrapolation tests of 41 species, with color scale indicating the MAE values shown to the right. The average MAE of 5 tests on the target element with different random seeds is written below the name of the corresponding element in the periodic table, while the MAE is denoted as "None" if the element does not appear in the dataset. **(b-d)** Average test MAEs of transfer learning on N, Re and Ru, in which 2, 5, 10, 20, 30 samples are added into the training sets, respectively. All points in the figure are the average of 5 tests with different samples included in the training process. The lines are drawn to exhibit the change of MAEs.

## Discussion

The contribution of MDNs in MT-MD DimeNet++ on its improved performance is explored by comparing MT-MD DimeNet++ to multi-task MLP (MT-MLP) DimeNet++, in which the MDN layers at the end of MT-MD DimeNet++ are substituted with MLPs. The test results of different models for the MT I2I task described in **Fig 1d** is shown in **Table 2**, and all of the results are again the average of 10 tests. MDNs do have an advantage over MLPs for the structure relaxation prediction task, suggested by a better performance of MT-MD model than that of MT-MLP model of the same architecture. As a result, the assumption that the bond length change is more likely to follow a distribution rather than being deterministic values is justified.

Table 2. Comparison of different types of MT DimeNet++.

| Type of Model | Architecture | Test MAE AEVR. | Test MAE STD. | Test STDAE AEVR. | Test within 0.02 AVER. | Test within 0.1 AVER. | Run Time AVER. | Number of Parameters |
|---|---|---|---|---|---|---|---|---|
| MT-MLP DimeNet++ | MB3-SRB2 | 0.0889 | 0.0150 | 0.351 | 0.378 | 0.828 | 6773.94 | 4569735 |
| MT-MD DimeNet++ | MB3-SRB2 | 0.0875 | 0.0135 | 0.310 | 0.384 | 0.823 | 7258.57 | 4569855 |

The function of dopant features in the networks is further investigated in order to understand the essential role of dopants in the catalytic performance of SAAs. First, the significance of dopant features and adsorbed atom features in GBR models is studied by only retaining one of them as the structure information of the configuration. Compared with the original model using both information to represent structures, the results in **Table S11** show that the excellent performance of SF-based models is attributed to dopant features rather than adsorbed molecule properties. Accordingly, the conventional GNN algorithms are further modified by adding the updated node embeddings of dopants directly to the resulting graph-level crystal embeddings, in order to emphasize the impact of dopants. It turns out that such a simple adjustment in the networks can lead to an improvement of performance, as illustrated in **Fig. S2**. For DimeNet-based modes, the embeddings of incident edges of dopants are averaged to obtain the dopant node embeddings and added to all edge embeddings of corresponding crystal graphs in each IB. However, the test results do not show a notable enhancement from those of original models. A better approach to emphasize the dopants in the edge-based message-passing models remains to be explored.

In summary, we propose a novel form of GNN model named multi-task mixture density (MT-MD) DimeNet++ to predict $E_{CO}$ of Cu-based SAAs, in order to optimize SAA catalysts for $CO_2$RR. In the model, empowered by MDNs, an SRB is applied to predict the change of atom positions after optimization, and the performance on $E_{CO}$ prediction task has improved notably when the relaxed structure information is encoded into the weight parameters during training. The model also exhibits extraordinary capacity on surface extrapolation tasks, so that $E_{CO}$ of SAAs constructed on high index Cu surfaces can be predicted accurately using the model only trained on low index data. Transfer learning to unobserved doping species is also feasible with only 5-10 extra SAAs samples doped by the test element provided, which can save a large amount of computational costs. Further experiments have been conducted to demonstrate the contribution of MDNs on the improved performance. It can be expected the presented framework can be applied with other advanced GNN models to predict adsorption energy of different intermediates to tackle the catalyst design challenge.

## Methods

**Details of Density function theory (DFT) computation for building the SAA dataset and training of ML models**

The DFT calculations were performed using Vienna *ab initio* simulation package (VASP) [76], with the project-augmented wave potential and a 400 eV cut-off energy [77]. Revised Perdew–Burke–Ernzerhof (rPBE) [78, 79] was used to describe exchange and correlation interaction within the generalized gradient approximation (GGA). For magnetic systems, spin-polarized computations were carried out. A (3 × 3 × 1) Monkhorst-Pack mesh for k-point sampling was performed for Brillouin-zone integration, and the smallest space between two k-points was set as 0.3 Å$^{-1}$. The vacuum distances were set larger than 20 Å, and the Cu atoms at the bottom two layers were frozen during relaxation, until the convergence threshold, i.e., $1 \times 10^{-4}$ eV/atom for energy and 0.05 eV/atom for force, was reached. After that, $E_{CO}$ was derived from energies of relaxed systems by subtracting the energies of CO and bare SAA from the energy of SAA adsorbed with a CO molecule. The *ab initio* molecular dynamics (AIMD) was done at 300K with an NVT ensemble using a Nosé-Hoover thermostat [80, 81] for 10 ps with a 1.0 fs time step.

All of the GNN models were trained under the framework implemented by OCP [64] to make a fair comparison, and the GBR algorithm was applied using the Scikit-learn package [82]. All of the GNN models in this work, including MT ones, were trained on a NVIDIA T4 GPU, and the GBR based models were done on an Intel i7-8700 CPU core.

**Directional message passing in DimeNet and DimeNet++**

When updating the directional embedding of a single edge in DimeNet, two pieces of information are incorporated: the edge encoding from the previous layer and the directional encoding aggregated from its incident edges. The latter includes angle data between the two edges, as well as edge length data from the incident edge. To be more precise, the directional message-passing process in the $(l + 1)_{th}$ layer can be formulated as

$$m_{ji}^{(l+1)} = U_t\left(m_{ji}^{(l)}, \sum_{k \in N_j} M_t\left(m_{kj}^{(l)}, e_{ji}, a_{(kj,ji)}\right)\right), \tag{1}$$

in which $m_{ji}$ denotes the directional edge embedding sent from atom $j$ to $i$, and $N_j$ denotes neighbor atoms of $j$ within the cut-off radius. $U_t$ and $M_t$ are the update function and message-passing function, respectively, as defined in [35]. $e_{ji}$ is the expansion of distance between $j$ and $i$ ($d_{ji}$) using radius basis, and the angle between bond $ji$ and $kj$, as well as $d_{kj}$, are incorporated into $a_{(kj,ji)}$ using spherical harmonics and spherical Bessel functions of the first and the second kind, establishing 2D spherical Fourier-Bessel basis, which achieves equivariance with respect to actions in the SO(3) group [83-87]. As the core part of the operation, the message-passing function can be expressed as

$$Msg = \sum_{k \in N_j} (W_a * a_{(kj,ji)}) * W_{bilinear} * [(W_e * e_{ji}) \odot \sigma_m(W_m * m_{kj}^l + b_m)], \tag{2}$$

where $W$, $b$ and $\sigma$ represent weight, bias and the activation function in fully connected layers, respectively, and $\odot$ represents element-wise multiplication. After that, the derived message is to be combined with $m_{ji}^{(l)}$ and several Residual layers [88] are applied to improve stability of the network and prevent gradient vanishing during training.

Noting that the bilinear layer is computationally costly, DimeNet++ replaces it with a simple Hadamard product with MLPs to accelerate the computation as

$$Msg = \sigma_{up}\left(W_{up}\left[\sum_{k \in N_j} W_{a2} * (W_{a1} * a_{(kj,ji)}) \odot \sigma_{down}\left(W_{down}\left(W_{e2} * (W_{e1} * e_{ji}) \odot \right.\right.\right.\right.$$
$$\left.\left.\left.\left.\sigma_m(W_m * m_{kj}^l + b_m)\right)\right)\right]\right). \tag{3}$$

With additional model architecture adjustments, DimeNet++ can reach an 8x quicker performance than DimeNet.

**Mixture density networks (MDNs)**

MDNs assume that the labels follow an unknown conditional distribution given the samples and use a set of Gaussians to approximate the distribution. The outputs of MDNs are not the label values, but the means, standard deviations and weight coefficients of Gaussians. The loss function of MDNs can be written as

$$\mathcal{L}_{MDN} = -\log \sum_{n=1}^{N} p_n * \mathcal{N}_n(y_{true}|\mu_n, \sigma_n), \tag{4}$$

in which $μ_n$ and $σ_n$ represent the mean and standard deviation of the $n_{th}$ Gaussian distribution $\mathcal{N}_n$, respectively, and $p_n$ denotes the contribution of the Gaussian to the real distribution. During training, the value of $\mathcal{L}_{\mathcal{MDN}}$ is reduced, so that the labels $y_{true}$ have a larger and larger probability to appear in the derived mixed distribution, which therefore characterizes the system.

**Cross-stitch units (CSUs)**

Assuming that $f_{I,1}$ and $f_{I,2}$ in the form

$$f_{I,1} = \begin{pmatrix} x_{I,1}^1 & \cdots & x_{I,1}^n \end{pmatrix}^T \tag{5}$$

$$f_{I,2} = \begin{pmatrix} x_{I,2}^1 & \cdots & x_{I,2}^n \end{pmatrix}^T \tag{6}$$

are the inputs of the CSU, i.e., the output edge embeddings of the two IBs above the CSU, the CSU should be a $2n*2n$ learnable matrix. The cross-stitch operation can be formulated as

$$f_{O,1} = \left( \sum_{i=1}^n x_{I,1}^i a_{i1} + \sum_{j=1}^n x_{I,2}^j a_{j+n,1} \quad \cdots \quad \sum_{i=1}^n x_{I,1}^i a_{in} + \sum_{j=1}^n x_{I,2}^j a_{j+n,n} \right)^T \tag{7}$$

$$f_{O,2} = \left( \sum_{i=1}^n x_{I,1}^i a_{i,n+1} + \sum_{j=1}^n x_{I,2}^j a_{j+n,n+1} \quad \cdots \quad \sum_{i=1}^n x_{I,1}^i a_{i,2n} + \sum_{j=1}^n x_{I,2}^j a_{j+n,2n} \right)^T \tag{8}$$

in which $a_{ij}$ denotes the element in the $i_{th}$ row and $j_{th}$ column in a CSU matrix, while $f_{O,1}$ and $f_{O,2}$ are outputs of the CSU, as well as the inputs of the two IBs below it. It can be seen that $f_{I,1} = f_{O,1}$ if the CSU is an identity, and otherwise the CSU can model how to share the two sets of weight parameters in two connected IBs by searching for appropriate $a_{ij}$ values to control the contribution of each element of two input features on the output, therefore deciding the effective MT network architecture.